\documentclass{emulateapj}
\newcommand{\be}{\begin{equation}}
\newcommand{\ba}{\begin{eqnarray}}
\newcommand{\ee}{\end{equation}}
\newcommand{\ea}{\end{eqnarray}}
\newcommand{\etal}{et al.\ }
\newcommand{\etalb}{et al.}
\newcommand{\nb}{\bar{n}}
\newcommand{\Ni}{N_{\rm ion}}
\newcommand{\Omm}{\Omega_m}
\newcommand{\HI}{\rm H\,I}

\newcommand{\Lya}{\mbox{Ly$\alpha$} }
\newcommand{\Ly}{\mbox{Ly$\alpha$}}
\newcommand{\Oml}{{\Omega_{\Lambda}}}

\usepackage{apjfonts}

\begin{document}
\title{Gamma-Ray Bursts versus Quasars: Lyman-$\alpha$ Signatures of 
Reionization versus Cosmological Infall}

\author{Rennan Barkana}
\affil{School of Physics and Astronomy, The Raymond and Beverly Sackler
Faculty of Exact Sciences, \\
Tel Aviv University, Tel Aviv 69978, ISRAEL; barkana@wise.tau.ac.il}

\author{Abraham Loeb\altaffilmark{1}} 
\affil{Institute for Advanced Study, Princeton, NJ 08540; 
aloeb@cfa.harvard.edu}

\altaffiltext{1}{Guggenheim Fellow; on sabbatical leave from the
Astronomy Department, Harvard University, Cambridge, MA 02138}

\begin{abstract}
Lyman-$\alpha$ absorption is a prominent cosmological tool for probing
both galactic halos and the intergalactic medium at high redshift. We
consider a variety of sources that can be used as the Lyman-$\alpha$
emitters for this purpose. Among these sources, we argue that quasars
are the best probes of the evolution of massive halos, while
$\gamma$-ray bursts represent the cleanest sources for studying the
reionization of the intergalactic medium.
\end{abstract}


\section{Introduction}

Over the past three decades, quasars have been traditionally used to
probe the intervening intergalactic medium (IGM) at redshifts $z\la 6$
when this medium was highly ionized. The polarization anisotropies of
the microwave background measured by WMAP indicate that the IGM was
significantly ionized at a much higher redshift $z\sim 17\pm 4$
\citep{WMAP}, possibly implying two separate reionization phases
separated by a prolonged period of nearly complete recombination
\citep{WL03,WL03b,C03}. Attention is now shifting towards direct
observations of the epoch of reionization. The process of reionization
started with the emergence of individual \ion{H}{2} regions centered
on discrete sources, which eventually overlapped [see reviews by
\citet{review1, review2}]. Observationally, a central goal is to
determine the average neutral fraction of the IGM, and possibly even
map the topology of the \ion{H}{2} islands, as a function of
redshift. The main problem with using resonant Ly$\alpha$ absorption
by neutral hydrogen for this purpose is that the corresponding
resonant cross-section is too large; a neutral fraction as small as
$\sim 10^{-3}$ is already sufficient to suppress the quasar flux below
detectability. For example, the spectrum of the Sloan Digital Sky
Survey (SDSS) quasar at $z= 6.28$ shows complete absorption [the
so-called \citet{GP} trough] of its spectral flux shortward of the
Ly$\alpha$ resonance wavelength at the quasar redshift \citep{z6p3},
but it remains unclear whether the IGM is highly ionized or whether it
contains pockets of neutral gas at this redshift \citep{me02,fan02}.

The use of other atomic lines is problematic for other reasons. For
example, absorption by metal lines \citep{oh02} can only probe regions
that had been previously metal-enriched by supernovae and the
ionization state of this disturbed gas may not be representative of
the entire IGM (since most metals may reside within \ion{H}{2} bubbles
near their source galaxies). Another possible probe, 21 cm absorption
by neutral hydrogen, suffers from a cross-section that is too {\it
small} (the opposite problem from Ly$\alpha$) leading to a very low
optical depth and requiring the use of very bright radio sources ($\ga
10$ mJy) that should be rare in the early universe
\citep{cgo02,fl02}. Detection of 21 cm absorption or emission against
the microwave background is technologically challenging \citep{tozzi,
iliev, furla} and is likely to be overwhelmed by foregrounds
\citep{matteo, loeb}.

It is therefore important to explore any possible way of using the
Ly$\alpha$ absorption signature near the source in order to measure
the IGM neutral fraction. In fact, for photon wavelengths longer than
the Ly$\alpha$ resonance at the source redshift, the absorption
cross-section is weaker (since these photons are offset from
resonance) and so the quasar spectrum should gradually recover its
unabsorbed spectral flux level as the wavelength becomes redder. The
detailed spectral shape of the red damping wing of the Gunn-Peterson
trough can potentially be used to infer the neutral hydrogen fraction
in the IGM during the epoch of reionization. \citet{jordi98}
calculated this shape for the idealized case of a weak point source
embedded in a neutral uniformly expanding IGM. In reality, however,
the ionizing effect of a quasar on the surrounding IGM [the so-called
"proximity effect" \citep{prox1}] complicates this interpretation
\citep{CH00, MR00}. In addition, cosmological infall changes the IGM
density and velocity field near bright early quasars because these are
likely to be hosted by a massive galaxy; \citet{nature} have shown
that this effect can be used as an important tool in itself, as a way
to measure directly the masses of the host dark matter halos of
quasars, and to test fundamental aspects of the processes of galaxy
and quasar formation. Indeed, we explore this tool further in this
paper and consider how infall around quasars can be probed over a
broad range of halo masses and redshifts.

Given that quasars and their massive hosts perturb the IGM around them
both radiatively and gravitationally, the following question arises:
{\it Are there better sources than quasars for probing the Ly$\alpha$
damping wing during reionization?} Our answer is positive. These
sources are the afterglows of Gamma-Ray Bursts (GRBs), suggested
before for this purpose \citep{jordi98,loeb01} but with additional
advantages that we highlight in this paper.

GRBs are the brightest electromagnetic explosions in the universe, and
should be detectable out to redshifts $z>10$ \citep{LR00, Ciardi}.
High-redshift GRBs can be easily identified through infrared
photometry, based on the Ly$\alpha$ break induced by absorption of
their spectrum at wavelengths below $1.216\, \mu {\rm m}\,
[(1+z)/10]$. Follow-up spectroscopy of high-redshift candidates can
then be performed on a 10-meter-class telescope. There are three main
advantages of GRBs relative to quasars:

\begin{itemize}

\item The afterglow flux at a given observed time lag after the
$\gamma$-ray trigger is not expected to fade significantly with
increasing redshift, since higher redshifts translate to earlier times
in the source frame, during which the afterglow is intrinsically
brighter \citep{LR00, Ciardi}. For standard afterglow lightcurves and
spectra, the increase in the luminosity distance with redshift is
compensated by this cosmic time-stretching effect.

\item In the standard $\Lambda$CDM cosmology, galaxies form
hierarchically, starting from small masses and increasing their
average mass with cosmic time. Hence, the characteristic mass of
quasar black holes and the total stellar mass of a galaxy were smaller
at higher redshifts, making these sources intrinsically fainter
\citep{Wyithe}. However, GRBs are believed to originate from a stellar
mass progenitor and so the intrinsic luminosity of their engine should
not depend on the mass of their host galaxy. At high redshifts, a GRB
should strongly dominate the luminosity of its host galaxy. Hence GRBs
are expected to be increasingly brighter than any other competing
source as higher redshifts are considered.

\item Since the GRB progenitors are believed to be stellar, they likely 
originate in the most common star-forming galaxies at a given redshift
rather than in the most massive host galaxies, as is the case for
bright quasars. Low mass host galaxies induce only a weak ionization
effect on the surrounding IGM and do not greatly perturb the Hubble
flow around them. Hence, the Ly$\alpha$ damping wing should be closer
to the idealized unperturbed IGM case \citep{jordi98} and its detailed
spectral shape should be easier to interpret. Note also that unlike
the case of a quasar, a GRB afterglow can itself ionize at most $\sim
4\times 10^4 E_{51} M_\odot$ of hydrogen if its UV energy is $E_{51}$
in units of $10^{51}$ ergs, and so it should have a negligible cosmic
effect on the surrounding IGM.

\end{itemize}

Although the nature of the central engine that powers the relativistic
jets of GRBs is unknown, recent evidence indicates that GRBs trace the
formation of massive stars \citep{bloom,kul,Tot97,Wij98,BlNat00}.
Since the first stars are predicted to be predominantly massive
\citep{Abel,BCL2002}, their death might give rise to large numbers of
GRBs at high redshifts. Detection of high-redshift GRBs could probe
the earliest epochs of star formation, one massive star at a time. The
upcoming {\it Swift} satellite (see http://swift.gsfc.nasa.gov/),
scheduled for launch in May of 2004, is expected to detect about a
hundred GRBs per year. \citet{BrL2002} calculated the expected
redshift distribution of GRBs; under the assumption that the GRB rate
is simply proportional to the star formation rate they found that
about a quarter of all GRBs detected by {\it Swift} should originate
at $z > 5$ \citep[see also][]{LR00}. This estimate is rather uncertain
because of the poorly determined GRB luminosity function.  We caution
further that in principle, the rate of high-redshift GRBs may be
significantly suppressed if the early massive stars fail to launch a
relativistic outflow. This is possible, since metal-free stars may
experience negligible mass loss before exploding as a supernova. They
would then retain their massive hydrogen envelope, and any
relativistic jet might be quenched before escaping the star
\citep{Heg03}. However, localized metal enrichment is expected to
occur rapidly (on a timescale much shorter than the age of the
then-young universe) due to starbursts in the first galaxies and so
even the second generation of star formation could occur in an
interstellar medium with a significant metal content, resulting in
massive stars that resemble more closely the counterparts of
low-redshift GRB progenitors.

In this paper we compare quantitatively the spectral profile of
Ly$\alpha$ absorption for quasars and GRB afterglows, including both
resonant absorption and the damping wing and taking into account the
ionization and infall effects induced by their host galaxies and halos
on the surrounding IGM. We present our detailed model in \S 2, and
then illustrate it in \S 3. In \S 4 we show how quasars can be used as
a tool for measuring infall, which probes the properties of massive
halos and their formation process. In \S 5 we show how, on the other
hand, GRBs can be used to cleanly probe the reionization of the
IGM. Finally, we give our conclusions in \S 6.

\section{Modeling \Lya transmission}

\subsection{Summary of Simple Models}

We assume throughout a standard $\Lambda$CDM cosmology. For the
contributions to the energy density, we assume ratios relative to the
critical density of $\Omm=0.27$, $\Oml=0.73$, and $\Omega_b=0.044$,
for matter, vacuum (cosmological constant), and baryons,
respectively. We also assume a Hubble constant $H_0=71\mbox{ km
s}^{-1}\mbox{ Mpc}^{-1}$, and a primordial nearly scale invariant
power spectrum (with a slowly varying spectral index) with
$\sigma_8=0.84$, where $\sigma_8$ is the root-mean-square amplitude of
mass fluctuations in spheres of radius $8\ h^{-1}$ Mpc. These
parameter values are based on recent temperature anisotropy
measurements of the cosmic microwave background in combination with
other large scale structure measurements, and the full set of
parameters is given by \citet{WMAP}.

An ionizing source embedded within the neutral IGM can ionize a region
of maximum proper size 
\be R_{\rm max} = 0.942\, \left( \frac{N_{\rm tot}} {10^{70}} \right)^
{1/3}\, \left( \frac{1+z} {8} \right)^{-1}\, \left( \frac{\Omega_b
h^2} {0.0222} \right)^{-1/3}\, {\rm Mpc}\ , \label{eq-Rmax} \ee
assuming that recombinations are negligible and that all $N_{\rm tot}$
ionizing photons are absorbed by hydrogen atoms. More generally, the
evolution of an \ion{H}{2} region, including a non-steady ionizing
source, recombinations, and cosmological expansion, is given by the
expanding Str\"{o}mgren sphere equation \citep{HIIgrowth},
\be \frac{dV}{dt}= \frac{1}{\nb_H^0} \frac{dN_{\gamma}}{dt}- \alpha_B 
\nb_H^0 V (1+z)^3\ , \label{HIIgrowth} \ee where $V$ is the comoving 
volume of the \ion{H}{2} region, $dN_{\gamma}/dt$ is the number of
photons per unit time output by the source, $\nb_H^0$ is the $z=0$
cosmic mean number density of hydrogen, and the case B recombination
coefficient for hydrogen at $T=10^4$ K is $\alpha_B=2.6\times
10^{-13}$ cm$^3$ s$^{-1}$

When dealing with \Lya emission and absorption, it is useful to
associate a redshift $z$ with a photon of observed wavelength
$\lambda_{\rm obs}$ according to $\lambda_{\rm obs}=
\lambda_{\alpha}(1+z)$, where the \Lya wavelength is $\lambda_{\alpha}=
1215.67 $\AA. In the absence of infall, the optical depth $\tau_{\rm
damp}$ due to the IGM damping wing of neutral gas between $z_1$ and
$z_2$ (where $z_1 < z_2 < z$) can be calculated analytically
\citep{jordi98}. For a source above the reionization redshift, $z_2$
is the redshift corresponding to the blue edge of the \ion{H}{2}
region, and $z_1$ is the reionization redshift. In the limit where
$z_2$ is very close to $z$ while $z_1$ is not, the full formula can be
approximated as
\ba \tau_{\rm damp} & & \approx \left\{ \frac{30.2\mbox{\AA}} 
{\Delta \lambda} - \frac{0.0140\times 8}{1+z}\, \log
\left[ \frac{1.95 \times 10^4 \mbox{\AA}} {\Delta \lambda}\, 
\frac{1+z}{8} \right] \right\} \nonumber \\ & & \times \left( 
\frac{1+z}{8} \right)^{5/2}\, \left( \frac{\Omm h^2}{0.136} 
\right)^{-1/2}\, \left( \frac{\Omega_b h^2} {0.0222} \right)\ , 
\label{eq-damp} \ea where $\Delta \lambda = \lambda_{\alpha}\, (z-z_2)$ 
and in this limit the result is independent of $z_1$.

In addition to the damping wing due to neutral gas that lies outside
the \ion{H}{2} region, photons may also be absorbed near the \Lya
resonance by the small neutral hydrogen fraction within the ionized
region. In the absence of clumping, this resonant absorption produces
an optical depth \be
\tau_{\rm res} = \tau_{\rm GP}\, x_{\HI}\, \Delta\ , 
\label{eq-res} \ee assuming gas at relative density $\Delta = 
\rho_g /\bar{\rho}_g$ and a neutral fraction $x_{\HI}$, where the 
standard \citep{GP} optical depth is
\be
\tau_{\rm GP}=4.83\times 10^5 \left( {1+z\over 8} \right)^{3 / 2} 
\left({\Omm h^2\over 0.136}\right)^{-{1 / 2}} \left({\Omega_b h^2
\over 0.0222} \right)\ . \label{G-P} \ee
The neutral fraction is given by ionization equilibrium. The optical
depth at a proper distance $R$ from a source that spews
hydrogen-ionizing photons into the IGM at a rate $dN_{\gamma}/dt$ is
\ba \tau_{\rm res} & & = 0.731\, \Delta^2\, \left( \frac{R} {1\, 
{\rm Mpc}} \right)^2\, \left( \frac{dN_{\gamma}/dt} {10^{57}\, {\rm
s}^{-1}} \right)^{-1}\, \left( \frac{\bar{\sigma}_H} {2 \times
10^{-18}\, {\rm cm}^2} \right)^{-1} \nonumber \\ & & \times \left(
\frac{1+z} {8} \right)^{9/2}\, \left( \frac{\Omm h^2} {0.136}
\right)^{-1/2}\, \left( \frac{\Omega_b h^2} {0.0222} \right)^2\ , 
\label{eq:res} \ea where $\bar{\sigma}_H$ is the frequency-averaged 
photoionization cross section of hydrogen (which depends on the
ionizing photon spectrum), and this result uses the value of
$\alpha_B$ given above.

Previous studies that included the absorption of gas in the \ion{H}{2}
region \citep{CH00,MR00} have all assumed that the mean gas density in
this region is equal to the cosmic mean density, and for the most part
have treated gas clumping very simply, taking a constant clumping
factor $C \equiv <\Delta^2>/<\Delta>^2\, \sim 10$ to enhance both
$\tau_{\rm res}$ and the recombination rate during the growth of the
\ion{H}{2} region [see eq.~(\ref{HIIgrowth})]. More recently, 
\citet{HC02} used numerical simulations directly, while \citet{zoltan} 
assumed a lognormal distribution of gas densities. In the following
section we account for infall and also use a realistic
redshift-dependent distribution of gas clumping that is based on
numerical simulations.

\subsection{Gas Infall and Clumping}

\label{sec-infall}

In the standard CDM-dominated picture of galaxy formation, galaxies
form within dark matter halos that themselves form only in regions of
high initial density. Since the density field is generated according
to an initial power spectrum, the region around a halo is strongly
correlated with the region that leads to the halo itself and the
former also tends to have a higher than average initial density. As
the dense central region undergoes gravitational collapse, the
surrounding region is drawn in towards the forming halo. Thus, by the
time a given dark matter halo has finally virialized, a much larger
region of surrounding gas has already acquired a significant infall
velocity and, by mass conservation, a large overdensity. Infall
affects resonant \Lya absorption through $\Delta$
[eq.~(\ref{eq-res})] and also affects the damping wing, for which
the optical depth is no longer given by a simple formula such as
eq.~(\ref{eq-damp}) but must be numerically integrated over the infall
profile.

\citet{Infall} developed a model for calculating the initial density
profile around overdensities that later collapse to form virialized
halos. This model, which we adopt in this paper, accounts for the
so-called ``cloud-in-cloud'' problem: if the average initial density
on a certain scale is high enough to collapse by a given redshift,
this region may be contained within a larger region that is itself
also dense enough to collapse; in this case the original region should
be counted as belonging to the halo with mass corresponding to the
larger collapsed region. Technically, \citet{Infall} adopted the
definition of halos in terms of the initial density field as proposed
by \citet{bc91}, who improved on the original \citet{PS} model by
including the cloud-in-cloud problem in the definition and in the
statistics of halos.

\citet{Infall} found that the mean initial density profile around an
overdensity that leads to a halo depends on both the mass of the halo
and its formation redshift. At high redshift, low mass halos tend to
have a stronger density enhancement around them that extends out to
larger distances (in units of the initial radius containing the virial
mass of the eventual halo); this is a result of the presence of
stronger correlations on smaller scales in the standard cosmological
model. However, at lower redshifts, low mass halos tend to be
surrounded by initial regions of relatively low density. This can be
understood as follows: low mass halos that were initially surrounded
by a high overdensity would most likely end up as part of a larger
halo, since even large halos are common at low redshift; therefore,
isolated low mass halos at low redshift should typically be surrounded
by initial regions of {\it low}\/ density, preventing them from
becoming part of a larger collapsed halo.

Starting from this mean initial profile, spherical collapse can be
used to obtain the final density and velocity profile surrounding the
virialized halo. In reality, there will be some variation in the
initial profiles, with each leading to a different final profile. This
scatter was also studied by \citet{Infall}, but in this paper we adopt
the mean profile described above, which accounts for the overall
trends in the infall profile as a function of the halo parameters.

Once the dark matter infall profile is fixed, we assume that the
infalling gas follows the dark matter at large distances from the
halo, where pressure gradients are negligible compared to
gravitational forces (i.e., we only consider halos well above the
Jeans mass of the intergalactic gas). When the gas finally does fall
into the halo, incoming streams from all directions strike each other
at supersonic speeds, creating a strong shock wave. Three-dimensional
hydrodynamic simulations show that the most massive halos at any time
in the universe are indeed surrounded by strong, quasi-spherical
accretion shocks \citep{Abel, EliSPH}. These simulations show that
along different lines of sight, the shock radius lies at a distance
from the halo center of around 1--1.3 times the halo virial radius. We
adopt 1.15 as an illustrative value, noting that the results are not
altered substantially as long as the shock radius is fairly close to
the virial radius.

We calculate gas infall down to the radius of the accretion shock, and
neglect any \Lya absorption due to the post-shock gas. The post-shock
gas is heated to around the virial temperature which, in halos we
consider, is $\ga 10^4$ K. In halos where the temperature is far
greater than $10^4$ K the gas is highly collisionally-ionized in
addition to the effect of the intense radiation field near the source,
and the absorption should be weak. If the gas is instead shocked to a
temperature close to $10^4$ K then it will subsequently cool and
likely collapse onto the galactic disk. Hydrodynamic simulations are
required in order to check for the possibility of having a thin cold
shell of shocked gas that may produce some absorption, but in any case
such absorption will be seen well to the blue side of the absorption
due to the infalling pre-shock gas.

In addition to the increase in the mean gas density due to infall, the
gas is also clumped, and this has an additional effect on the
recombination rate and on \Lya absorption. For the distribution of gas
clumps we adopt the distribution that \citet{clumping} constructed
based on numerical simulations at $z=2$--4 and extrapolated to other
redshifts: \be P_V(\Delta_C) = A\, \exp\, \left[ -
\frac{(\Delta_C^{-2/3} - C_0)^2} {2 (2 \delta_0/3)^2}\right]\,
\Delta_C^{-\beta}\ , \label{clumping} \ee where clumps at an
overdensity $\Delta_C$ account for a volume fraction $P_V(\Delta_C)
d\Delta_C$ of the gas. Following \citet{clumping}, we adopt
$\beta=2.5$ for $z \ge 6$ and slightly lower values of $\beta$ at
lower redshifts, set $\delta_0 = 7.61/(1+z)$, and fix the other
parameters by requiring the total volume and mass fractions to be
normalized to unity. \citet{clumping} compared their clumping
distribution only to post-reionization simulations, and only up to
$\Delta_C \sim 100$; however, we are justified in adopting this
distribution (up to $\Delta_{\rm max} = 100$) since we apply it only
within \ion{H}{2} regions, where the gas has already been photoionized
and heated, and we also demonstrate below that all our observable
predictions are insensitive to the high-density tail of the
distribution. In our model we assume that eq.~(\ref{clumping}) applies
with $\Delta_C$ interpreted as the extra overdensity due to clumping,
measured relative to the mean gas density $\Delta$ (which itself may
be higher than the cosmic mean, due to infall).

\citet{clumping} modeled reionization with a model that assumes that
all clumps up to some critical $\Delta_C$ are fully ionized while
clumps above this $\Delta_C$ are completely neutral. This model is
inadequate for our purposes, since the small neutral fraction expected
in the low-density clumps is in fact crucial in determining the total
\Lya absorption, and in addition high-density clumps are in reality
not self-shielded and fully neutral except at very high densities that
do not affect our results. We thus assume that clumps are optically
thin instead, and find the neutral fraction $x_{\HI}(\Delta_C)$
separately for each $\Delta_C$ based on ionization equilibrium of gas
at that overdensity. This implies that the neutral fraction
essentially increases in proportion to $\Delta_C$ (except when
$x_{\HI}(\Delta_C) \rightarrow 1$, which occurs only for extremely
high overdensities and does not affect our results). Thus, we may
define one type of effective clumping factor in terms of the total
neutral fraction compared to the value it would have in the absence of
clumping: \be C_{x_{\HI}} = \int_0^{\Delta_{\rm max}} \frac{x_{\HI}
(\Delta \times \Delta_C)} {x_{\HI}(\Delta)}\, P_V(\Delta_C) \Delta_C\,
d\Delta_C\ , \label{eq-CxHI} \ee where $\Delta$ is the enhancement of
the mean density due to gas infall.

The recombination rate of a given gas element and the optical depth to
ionizing photons are both proportional to $C_{x_{\HI}}$. However, \Lya
absorption is not determined simply by the total number density of
hydrogen atoms, since it results from resonant absorption by separate
gas elements at a variety of overdensities; even if high-density
clumps produce complete absorption, photons at wavelengths that
resonate with gas in voids are strongly transmitted. Cosmological
volume is dominated by voids, and furthermore the mean fraction of a
given line of sight covered by gas of a given overdensity is simply
equal to the volume fraction occupied by gas at that overdensity,
regardless of the topology of the clumps [e.g., see footnote in \S 2.5
of \citet{minihalos}]. Thus, the effective clumping factor for \Lya
transmission $C_{{\rm Ly}\alpha}$ is determined by
\be e^{-\left[\tau_{\rm GP}\, C_{{\rm Ly}\alpha}\, \Delta\,  
x_{\HI}(\Delta) \right]} = \int_0^{\Delta_{\rm max}}\, e^{-\left[ 
\tau_{\rm GP}\, \Delta\, x_{\HI} (\Delta \times \Delta_C)\, \Delta_C 
\right]} P_V(\Delta_C)\, d\Delta_C\ . \label{eq:CLya} \ee
Note that when we use $C_{{\rm Ly}\alpha}$, we are in effect averaging
over the density fluctuations in the IGM. This could in principle be
compared directly to observations averaged over many different lines
of sight through the IGM. Any particular line of sight is expected to
show significant fluctuations (in the resonant absorption component)
relative to the average pattern that we predict. These fluctuations
are the high-redshift equivalent of the \Lya forest of absorption
features seen in the IGM at lower redshifts. Note that unlike resonant
absorption, damping wing absorption is determined by the cumulative
effect of a large column of gas (typically of order 1 Mpc long) and is
therefore unaffected by fluctuations and clumping that occur on much
smaller spatial scales.

The mean free path to ionizing photons depends on the total neutral
fraction and thus on $C_{x_{\HI}}$, since absorption of ionizing
photons is not a resonant process but is instead cumulative along a
line of sight. While $C_{{\rm Ly}\alpha}$ is typically dominated by
low values of $\Delta_C \la 1$, $C_{x_{\HI}}$ is sensitive to the high
integration limit $\Delta_{\rm max}$. However, this sensitivity is a
mathematical artifact, not a real physical dependence of the mean free
path of ionizing photons; it shows that if gas clumps were uniformly
distributed, so that the one-point distribution of gas density in
arbitrarily small volumes were given by eq.~(\ref{clumping}), then the
high-density tail of clumps would indeed dominate the value of
$C_{x_{\HI}}$. However, in reality gas clumps in the ionized IGM at
densities $\ga 20$ are found only around or inside localized, rare,
virialized halos. The clumping distribution is only correct as an
average over large volumes that contain a large statistical sample of
voids, filaments, and halos [but note that the simulations fitted by
\citet{clumping} could not probe a possible population of
minihalos]. An ionizing photon, for example, may travel a long
distance through a void before encountering any clump of significant
overdensity, and thus the mean free path of ionizing photons is
determined by the topology of the clumps and not just by the overall
one-point distribution of $\Delta_C$. \citet{clumping} solved this
problem approximately, finding a rough formula for the typical
distance $\bar{d}$ (defined as a mean free path) along a random line
of sight between encounters of clumps with $\Delta_C \ge
\widetilde{\Delta}_C$: \be
\bar{d}(\widetilde{\Delta}_C) \approx 60\,{\rm km\, s}^{-1}\,
H^{-1}(z)\, \left[ 1 - \int_{0}^{\widetilde{\Delta}_C}\,
P_V(\Delta_C) d\Delta_C\, \right]^{-2/3}\ , \label{mfp} \ee where
$H(z)$ is the Hubble constant at redshift $z$. If, for a given
$\widetilde{\Delta}_C$, we consider a line of sight of length $l$
through an \ion{H}{2} region, then the probability of encountering any
gas clump with $\Delta_C \ge \widetilde{\Delta}_C$ somewhere along
this line of sight is \be P_{\rm any}(> \widetilde{\Delta}_C) =
1-e^{-l/\bar{d} (\widetilde{\Delta}_C)}\ . \ee In this paper, we
consider absorption along relatively short lines of sight from a
source. Most such lines of sight will not contain any clumps with very
high density $\Delta_C$. Thus, we fix our maximum $\Delta_{\rm max}$
so that a line of sight of the length we are considering is reasonably
likely to go through at least one clump of that high a density. Since
\citet{clumping} compared eq.~(\ref{mfp}) to simulations averaged over
the mean IGM, in regions overdense due to infall we replace the line
of sight length $l$ by an effective length weighted by the overdensity
$\Delta$. In our model we set $\Delta_{\rm max}$ according to $P_{\rm
any}(> \Delta_{\rm max}) = 0.5$ (i.e., a typical line of sight), but
the absorption profiles are only weakly sensitive to the adopted
probability. The resulting values of $\Delta_{\rm max}$ range from
$\sim$1 for the smallest ($\sim 0.1$ Mpc) \ion{H}{2} regions up to
$\sim$ 50 for the largest ($\sim 10$ Mpc) \ion{H}{2} regions.

\subsection{Basic Halo Parameters}

\label{sec-IMFs}

The absorption profile due to \ion{H}{1} in the IGM depends on several
basic parameters. The first is the source redshift $z_S$. The second
is the source halo mass $M$, which determines the density and velocity
profile of infalling gas, as well as the virial radius and thus the
shock radius. The remaining three describe the source ionizing photon
production. One parameter is $dN_{\gamma}/dt$, the total rate at which
hydrogen ionizing photons from the source enter the IGM. The second
parameter is the age of the source $t_S$, which is the period of time
during which the source has been active (with an assumed constant
$dN_{\gamma}/dt$, for simplicity). Note that the source lifetime is
only important prior to the end of reionization, when the lifetime
affects the size of the surrounding \ion{H}{2} region. The final
parameter is $\bar{\sigma}_H$, the frequency-averaged photoionization
cross section of hydrogen, which affects the absorption and depends on
the spectrum of the ionizing photons.

Bright quasars are powerful ionizing sources and so the stellar
emission of their galactic host can be neglected. In order to predict
the ionizing intensity for a quasar with a given continuum flux, we
assume a typical quasar continuum spectrum as follows. We adopt a
power-law shape of $F_{\nu} \propto \nu^{-0.44}$ in the rest-frame
range 1190--5000\AA\ based on the SDSS composite spectrum
\citep{SDSScomp}, and $F_{\nu} \propto \nu^{-1.57}$ at 500--1190\AA\
using the composite quasar spectrum from the {\it Hubble Space
Telescope} \citep{EUVcomp}. Based on observations in soft X-rays
\citep{Xcomp}, we extend this power-law towards short wavelengths. We
assume that the brightest quasars shine at their Eddington luminosity
(but we also compare to the case of a tenth of Eddington), and we note
that for the SDSS composite spectrum
\citep{SDSScomp}, the total luminosity above 1190\AA\ equals 1.6 times
the total continuum luminosity at 1190--5000\AA. Thus, ionizing
photons stream out of the quasar's galactic host at the rate $\dot{N}
= 1.04\times 10^{56} M_8$ s$^{-1}$, where $M_8$ is the black hole
mass $M_{\rm BH}$ in units of $10^8M_\odot$. With these assumptions,
the black hole mass can be inferred from the observed continuum at
1350\AA with the conversion $F_{\nu} = 1.74 \times 10^{30} M_8$ erg
s$^{-1}$ Hz$^{-1}$.

In order to predict the \Lya absorption around a quasar we must
estimate the mass of its host halo. A tight correlation has been
measured in local galaxies between the mass of the central black hole
and the bulge velocity dispersion \citep{BHlocal1,BHlocal2}. This
relation also fits all existing data on the luminosity function of
high-redshift quasars within a simple model \citep{Wyithe,Wy03} in
which quasar emission is assumed to be triggered by mergers during
hierarchical galaxy formation. We use the best-fit \citep{BHlocal1,
BHlocal2, Wy03} relation, in which the black hole mass $M_{\rm BH}$ is
related to the circular velocity $V_c$ at the halo virial radius by
\be 
M_{\rm BH} = 8.8 \times 10^7 \left( \frac{V_c}{300\mbox{ km s}^{-1}} 
\right)^5 M_{\odot}\ . \ee
In general, we assume a source age $t_S \sim 10^7$--$10^8$ yr [see
\citep{Wy03} for a discussion].

When we consider GRB afterglows at redshifts $z\ga 7$, which are
typically found in relatively small galaxies, the above scalings imply
that the contribution of a central black hole in these galaxies is
negligible compared to the stars, and we must therefore include the
stellar output of ionizing radiation. For stars, we find it useful to
express $dN_{\gamma}/dt$, for given values of $M$ and $t_S$, as
follows: \be \frac{dN_{\gamma}}{dt} = \frac{M}{m_p}\,
\frac{\Omega_b}{\Omm}\, \frac{\Ni}{t_S}\ , \label{eq-dNdt} \ee where 
$m_p$ is the proton mass, and $\Ni$ gives the overall number of
ionizing photons per baryon. To derive the expected range of possible
values of $\Ni$, we note that it is determined by \be \Ni = N_{\gamma}
\, f_*\, f_{\rm esc}\ , \ee where we assume that baryons are incorporated
into stars with an efficiency of $f_*$, that $N_{\gamma}$ ionizing
photons are produced per baryon in stars, and that the escape fraction
(out to the virialization shock radius) for the resulting ionizing
radiation is $f_{\rm esc}$. We adopt $f_*=10\%$ (based on a rough
comparison of models to the observed cosmic star formation rate at low
redshift), and consider $f_{\rm esc}$ in the range 10$\%$--90$\%$.

The remaining variable $N_{\gamma}$ depends on the stellar initial
mass function (IMF). We consider two examples of possible IMFs. The
first is the locally-measured ``normal'' IMF of \citet{scalo}, along
with a metallicity equal to $1/20$ of the solar value, which yields
[using \citet{Leith99}] $N_{\gamma}=4300$. The second is an extreme
Pop III IMF, assumed to consist entirely of zero metallicity $M \ga
100\, M_{\odot}$ stars, which yields $N_{\gamma}=44000$ based on the
ionization rate per star, the stellar spectrum, and the main-sequence
lifetime of these stars \citep{Bromm}. In general, though, the
ionization state of hydrogen depends only on the ionizing photons that
are actually absorbed by hydrogen; therefore, the effective
$N_{\gamma}$ for hydrogen, as well as the effective cross-section
$\bar{\sigma}_H$ [see eq.~(\ref{eq:res})], are both affected by the
presence of helium, as explained in the following section.

\subsection{The State of Helium During Hydrogen Reionization}

Along with the \ion{H}{2} ionization front, we calculate the evolution
of the \ion{He}{2} and \ion{He}{3} fronts with equations analogous to
eq.~(\ref{HIIgrowth}). Initially, when the IGM is neutral, hydrogen
atoms absorb all 13.6--24.59 eV photons and helium atoms absorb all
$E>24.59$ eV photons. Since the number density of helium is only
$f_{He}= 7.9\%$ that of hydrogen (assuming $24\%$ helium by mass), the
\ion{He}{2} front easily keeps up with the expanding \ion{H}{2} front,
since most realistic sources produce $E>24.59$ eV photons relative to
13.6--24.59 eV photons in a much larger ratio than $f_{He}$. However,
the \ion{He}{3} front may either lag behind (for sources which emit
relatively few of the $E > 54.42$ eV photons that are needed to fully
ionize helium) or may catch up with the other two fronts.

Our calculations of the sizes of the various regions assume the same
infall overdensity and clumping factor for helium and for hydrogen,
and numerically solve for the expanding Str\"{o}mgren
spheres. However, for most sources the effect of helium can be
accurately described by one of two limiting cases. In the first case
(applicable to normal stars), the \ion{He}{3} front lags far behind
and covers a negligible volume compared to the common
\ion{H}{2} -- \ion{He}{2} front. In this case, all helium atoms within
the \ion{H}{2} region are ionized exactly once, and therefore the
effective $N_{\gamma}^{\rm rad}$ for determining the radius of the
\ion{H}{2} region is equal to the total number of ionizing photons
times $1/(1+f_{He})$. In this case as well, all $E > 54.42$ eV photons
are absorbed by \ion{He}{2} when they reach the \ion{He}{3} front, so
the effective $\bar{\sigma}_H$ for hydrogen as well as the effective
$N_{\gamma}^{\rm ion}$ for determining the ionizing intensity are both
determined by the spectrum and number of photons at $E < 54.42$ eV;
however, since this case is by definition accurate only for sources
which emit very few photons above 54.42 eV, the spectrum and intensity
are essentially equal to the result obtained from all $E > 13.6$ eV
photons emitted by the source.

In the second case (applicable to Pop III stars as well as to
quasars), the \ion{He}{3} front catches up with the
\ion{He}{2} front and creates a common \ion{H}{2} -- \ion{He}{3} front
beyond which both helium and hydrogen are fully neutral. In this case,
all helium atoms within the \ion{H}{2} region are ionized exactly
twice, and therefore the effective $N_{\gamma}^{\rm rad}$ is equal to
the total number of ionizing photons times $1/(1+2 f_{He})$. In this
case, all $E > 13.6$ eV photons emitted by the source reach the gas
within the \ion{H}{2} region and determine both $\bar{\sigma}_H$ and
$N_{\gamma}^{\rm ion}$. Although our numerical results below account
for the full complexity, where we now give representative numbers for
various sources we assume one of the two limiting cases above, as
appropriate for each type of source.

Thus, we find the following cross-sections for the sources discussed
in the previous subsection: $\bar{\sigma}_H = 2.8
\times 10^{-18}$ cm$^2$ for stars with a normal IMF, 
$\bar{\sigma}_H = 1.2 \times 10^{-18}$ cm$^2$ for stars with a Pop III
IMF, and $\bar{\sigma}_H = 2.3 \times 10^{-18}$ cm$^2$ for quasars.
Note that we have neglected some other radiative transfer effects; for
example, the degraded photons that are re-emitted by helium may ionize
hydrogen atoms as well, but only a minor fraction of the helium atoms
are expected to recombine on the short source timescale $t_S$.

The source ionizing spectrum and the presence of helium also affect
the gas temperature, which depends on photoheating as well as
adiabatic cooling, Compton cooling, and various atomic cooling
processes. \citet{HeRadTr} showed that, beyond heating and cooling
processes, the temperature profile inside an \ion{H}{2} region depends
on subtle radiative transfer effects; guided by their results, we
simply adopt characteristic temperatures of 40,000 K within
freshly-reionized \ion{He}{3} regions and 15,000 K in regions where
helium is not fully ionized. These temperatures affect the case B
recombination coefficients of the various atomic species [we adopt
values from \citet{atomic}], which in turn affect both the growth of
the various ionized regions and the final $\tau_{\rm res}$ at each
distance. The final effect of helium that we include as well is the
enhancement of the recombination rate due to the extra electrons from
ionized helium, where the enhancement factor equals $1+f_{He}$ in
\ion{He}{2} regions and $1+2 f_{He}$ in \ion{He}{3} regions.

\subsection{Transmission After the End of Reionization}

Once reionization of the universe is complete, the mean free path of
ionizing photons begins to rise. A UV background is established which
gradually becomes more uniform as each point in the IGM sees an
increasing number of ionizing sources. In this post-reionization era,
we assume that the mean free path of ionizing photons is much longer
than the line of sight that we consider, and thus it is unnecessary to
estimate the mean free path as was done in \S \ref{sec-infall}. After
reionization, there is also no limit to the \ion{H}{2} region
surrounding a source. In this case, there would still be a limit to
the \ion{He}{3} region around a quasar, if the cosmic UV background is
not sufficiently hard to fully ionize helium throughout the
universe. We neglect this, however, since the observable effect on
\Lya absorption would be minor, given that calculations with radiative
transfer show that the double ionization of helium produces a
characteristic gas temperature of only $\sim 1.5\times 10^4$ K in
regions that had already been reionized by a softer ionizing
background \citep{HeRadTr}.

In our post-reionization calculations, we add the mean UV background
to the ionizing intensity of the central source. This background
determines the average transmission level at large distances from the
particular galaxy being considered. We set the background level at
different redshifts so that our assumed distribution of gas clumping
matches observations; specifically, we use eq.~(\ref{eq:CLya}) with
$\Delta=1$ and $\Delta_{\rm max} = 100$ and compare it to the mean
transmission in the \Lya forest at various redshifts; we use
\citet{bernardi} for $z=2.5$--4 and \citet{songaila} for $z=4.5$--6,
we assume no UV background at $z>6.5$, and we smoothly interpolate in
the redshift intervals 4--4.5 and 6--6.5. Note that if much higher
redshifts ($z \ga 15$) are considered, the contribution of a cosmic UV
background might have to be considered around the time of a previous
reionization phase as suggested by the large-angle polarization of the
cosmic microwave background \citep{WMAP}. However, in the examples in
this paper we refer only to the final phase of reionization at $z \sim
7$, assuming that the universe had managed to recombine nearly
completely after a possible earlier reionization phase (see \S 1).

\subsection{Absorption by the Host Galaxy}
\label{sec-extinct}

The galactic host of any potential background light source may itself
absorb some flux near \Lya and thus muddle the interpretation of any
detected absorption, which is no longer exclusively due to \ion{H}{1}
in the IGM. The two main effects are absorption due to \ion{H}{1} gas
in the galactic host and extinction by dust.

As noted above, absorption due to the damping wing of \ion{H}{1} in
the IGM produces an optical depth that varies approximately inversely
with $\Delta \lambda$. However, since \ion{H}{1} gas in the galaxy is
concentrated in a compact region compared to the apparent distance
corresponding to the relevant range of $\Delta \lambda$, it produces
for photons with $\lambda_{\rm obs}=\lambda_{\alpha}(1+z)=
\lambda_{\alpha}\,(1+z_S)+ \Delta \lambda$ an optical depth
\be \tau_{\rm damp} = 7.26\, \left( \frac{N_{\rm H\, I}}
{10^{21}\,{\rm cm}^{-2}} \right)\, \left( \frac{1+z}{8} \right)^4\,
\left( \frac{1+z_S}{8} \right)^{-2}\, \left( \frac{\Delta
\lambda}{20\,\mbox{\AA}} \right)^{-2}\ , \ee
where $N_{\rm H\, I}$ is the total \ion{H}{1} column density in the
host galaxy at redshift $z_S$. Note that this formula does not assume
$|z-z_S| \ll 1$. The inverse-square dependence of $\tau_{\rm damp}$
due to the galactic host may in principle allow the isolation of this
effect from damped IGM absorption. However, a column density $\ga
10^{21}$ cm would obscure much of the region where damped absorption
could otherwise be measured, and could be partly degenerate with IGM
absorption if only a narrow wavelength window is available for
separating the two effects. As shown by \citet{jordi98}, such a
partial degeneracy may hamper an accurate measurement of the IGM
density through the normalization of $\tau_{\rm damp}$ due to the IGM;
however, as we show in \S \ref{sec-GRBs}, this degeneracy does not
preclude a definite {\it detection}\, of absorption due to a neutral
IGM, and it mostly disappears if the source redshift can be measured
accurately from additional absorption lines (other than Ly$\alpha$)
associated with the host galaxy. Note also that in cases where IGM
absorption is not complete in the \Lya blue wing, a neutral gas column
in the host galaxy could produce an observable, symmetric blue damping
wing.

The second source of nuisance absorption is extinction by dust in
the host galaxy. As a guide to the expected extinction we adopt the
mean Galactic extinction curve as modeled by \citet{extinction1}, and
normalize according to the dust-to-gas ratio of the diffuse Galactic
interstellar medium \citep{extinction2}. We find that extinction at
\Lya is higher than in the $V$ band according to
$A(\lambda_{\alpha})= 3.3\, A(V)$, where $A(\lambda)$ is the
extinction (in magnitudes) at rest-frame wavelength
$\lambda$. Although the extinction may make sources harder to detect,
the dependence on wavelength is expected to be smooth enough so as not
to distort the signature of \ion{H}{1} absorption. Specifically,
near the \Lya wavelength, a given neutral column produces extinction
\be A(\lambda) \approx 1.7\, \left[1-0.013\,\left( \frac{\lambda-
\lambda_{\alpha}} {10\,\mbox{\AA}} \right) \right]\, \left( 
\frac{N_{\rm H\, I}} {10^{21}\,{\rm cm}^{-2}} \right)\ ,
\label{eq-extinc} \ee
where all wavelengths in this expression are measured in the rest
frame. Note that this expression assumes the Galactic dust-to-gas
ratio, while a high-redshift galaxy is expected to have a lower mean
metallicity and dust content (although high metallicity may be found
in star-forming regions due to strong local metal enrichment).

\section{The Basic Model: Three Examples}

\label{sec-illust}

In this subsection we illustrate our predictions for several different
realistic sets of parameters. We consider a GRB and a quasar, each at
redshift 7 and having formed in a region that had not yet been
reionized by other sources (or had recombined after a possible earlier
reionization phase; see \S 1). The quasar is assumed to have an
observed flux $F_{\lambda}$ of $10^{-17}$ erg cm$^{-2}$ s$^{-1}$ \AA\
$^{-1}$ (typical of the SDSS high-redshift quasars) and to have been
radiating for $t_S=2 \times 10^7$ yr at the Eddington luminosity,
which implies in our model (see \S \ref{sec-IMFs}) a $1.7 \times 10^9
M_{\odot}$ black hole at the center of a halo of total mass $4.7
\times 10^{12} M_{\odot}$. In the other example, we assume that the
GRB goes off in a halo of mass $M=4 \times 10^8 M_{\odot}$ (typical of
$z=7$ GRBs; see \S \ref{sec-GRBs}), which hosts a galaxy of stars; we
consider either a normal IMF with $f_{\rm esc}=10\%$ and $t_S=1 \times
10^7$ yr, or a Pop III IMF with $f_{\rm esc}=90\%$ and $t_S=1 \times
10^8$ yr.

\begin{figure}[htbp]
\plotone{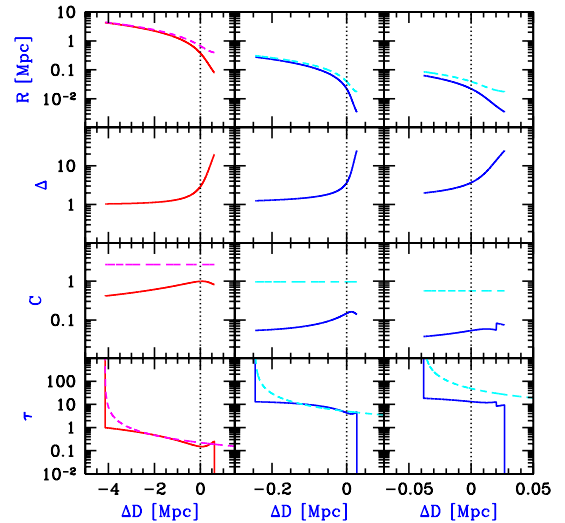} 
\caption{Properties of the gas in the \ion{H}{2} region versus
relative apparent position $\Delta D$. Panels correspond to a GRB host
with a normal stellar IMF (right panels) or a Pop III IMF (middle
panels), and to a quasar (left panels), all at redshift 7 in a region
that had not been pre-ionized (note the different $\Delta D$ ranges;
see text for the values of the input parameters). From bottom to top,
panels show the \Lya optical depth due to resonant absorption
($\tau_{\rm res}$; solid curves) and due to the damping wing
($\tau_{\rm damp}$; dashed curves); the clumping factors $C_{{\rm
Ly}\alpha}$ (solid curves) and $C_{x_{\HI}}$ (dashed curves); the gas
overdensity due to infall; and the actual distance of each gas element
from the ionizing source (solid curves), as well as the distance that
the same gas element would have reached in the presence of pure Hubble
expansion with no infall (dashed curves).}
\label{fig-clumps}
\end{figure}

Figure~\ref{fig-clumps} shows the resulting absorption profiles in
each of these cases, and also presents the physical properties of the
gas within the \ion{H}{2} region. A useful parameter is the apparent
line-of-sight position $\Delta D$ relative to the source redshift
$z_S$, derived directly from the observed wavelength by assuming
Hubble flow. The conversion, for a small wavelength difference $\Delta
\lambda$ relative to \Lya at $z_S$, is \be
\Delta D = 0.370\,\left( \frac {\Delta \lambda} {10\,\mbox{\AA}}
\right) \, \left( \frac{1+z_S}{8}\right)^{-5/2}\, \left(
\frac{\Omega_m h^2} {0.136} \right)^{-1/2}\,{\rm Mpc}\ . \ee The 
figure shows that the contributions to $\tau$ from the damping wing
($\tau_{\rm damp}$) and from resonant absorption ($\tau_{\rm res}$)
are comparable. Both become very large as the edge of the \ion{H}{2}
region is approached on the blue side (at $\Delta D < 0$). On the red
side, however, the smooth decline of $\tau_{\rm damp}$ with $\Delta D$
contrasts with the sharp drop in $\tau_{\rm res}$. Clearly,
regardless of the presence or absence of the damping wing, the
signature of infall identified by \citet{nature} is expected, i.e., a
sudden and significant drop in the transmitted flux at the wavelength
where resonant absorption starts to operate. Note that in order to
calculate the damping wing we assume a neutral IGM along the line of
sight down to $z=6.5$, although $\tau_{\rm damp}$ is very insensitive
to the precise lower limit [see eq.~(\ref{eq-damp})].

The figure shows the strong effect of infall, which produces resonant
absorption well into the red wing (i.e., at $\Delta D > 0$). In the
case of pure Hubble flow, $\tau_{\rm res}$ would drop to zero right at
$\Delta D=0$, and the drop would be gradual because of the increasing
ionizing intensity at small distances from the source. When infall is
included as part of a more realistic model, however, the infall
velocity results in absorption of photons at $\Delta z > 0$;
furthermore, the value of $\tau_{\rm res}$ does not show a gradual
drop since the high infall overdensity (and, therefore, high
recombination rate) compensates for the high ionizing intensity felt
by gas near the source. The sudden drop in $\tau_{\rm res}$ is due to
the inclusion of an accretion shock, which fully ionizes the gas and
cuts off absorption very close to the source. The inclusion of a
clumping distribution, and the dominance of voids in allowing
transmitted flux, leads to a strong reduction in $\tau_{\rm res}$
(i.e., $C_{{\rm Ly}\alpha}$ is significantly smaller than
$C_{x_{\HI}}$). In the quasar and Pop III star cases, helium is fully
ionized throughout the \ion{H}{2} region, but in the normal IMF case,
the GRB host galaxy produces only a small \ion{He}{3} region; the
optical depth jumps at this radius, due to the lower gas temperature
in the \ion{He}{2} region, and the transition can be seen at $\Delta D
= 0.02$ Mpc.

The effective clumping factor for transmission ($C_{{\rm Ly}\alpha}$)
varies over the range 0.03--1. On the other hand, the clumping factor
$C_{x_{\HI}}$ is always much larger, although its value ($\sim$0.5--3)
is still significantly lower than the values $\ga$ 10 that previous
studies have used which did not account for the rarity of high clump
densities along a given line of sight. Since it is defined in terms of
a ratio of neutral fractions [see eq.~(\ref{eq-CxHI})], $C_{x_{\HI}}$
is essentially independent of both $\Delta$ and the ionizing
intensity, and is simply a function of $\Delta_{\rm max}$. For the
cases shown in the figure, the proper radius of the \ion{H}{2} region
and the value of $\Delta_{\rm max}$ are respectively 0.063 Mpc and 1.8
(right panels), 0.28 Mpc and 3.7 (middle panels) and 4.3 Mpc and 31
(left panels). On the other hand, $C_{{\rm Ly}\alpha}$ is dominated by
low values of $\Delta_C$, is nearly independent of $\Delta_{\rm max}$,
and is essentially a function of the no-clumping ($C=1$) optical
depth, which itself is proportional to $\Delta^2 R^2$, where we used
the fact that the source ionizing intensity (which affects the neutral
hydrogen fraction) is inversely proportional to $R^2$. Indeed, when
the $C=1$ optical depth is large, the transmission in the voids (where
$\tau_{\rm res} \propto \Delta_C^2$ is low) becomes increasingly
dominant in the calculation of the overall mean transmission. Note
that our results differ quanitatively from the clumping model of
\citet{zoltan}, who found an effective clumping factor for transmission
of $\sim 0.04$ when $\tau_{\rm res}=1$, while we obtain $C_{{\rm
Ly}\alpha} \sim 0.4$ in this case.

The figure also shows that $\Delta \sim 20$ at the shock radius,
although the infall pattern depends on the halo mass and redshift; at
a given redshift, higher-mass halos correspond to rarer density peaks,
which tends to strengthen the infall pattern around them, but in
standard models there is less power on large scales and this works in
the opposite direction [see \citet{Infall}]. Gas with large $\Delta$
is located at a real radius $R$ that is much smaller than the radius
the same gas element would have reached in the absence of infall; this
is consistent with mass conservation. Observations of $\tau_{\rm
res}$ can in principle probe the density profile of cosmological gas
infall all the way from the shock radius out to the radius of the
\ion{H}{2} region, a range that represents $\sim 1.5$ orders of
magnitude of $R$. Note that for each gas element, the total velocity
relative to $z_S$ is $H(z_S)\, \Delta D$, while the Hubble velocity
component is $H(z_S)\, R$, where the Hubble constant at high redshift
is accurately approximated by \be H(z) = 835\, \left( \frac {1+z} {8}
\right)^{3/2}\, \left( \frac {\Omm h^2} {0.136} \right)^{1/2}\, {\rm\
km\ s^{-1}\ Mpc^{-1}}\ . \ee

The main conclusion from this figure is that a quasar affects a much
larger surrounding region than a typical GRB host, both in terms of
infall and ionization. Specifically, the region of significant infall
(measured by $\Delta > 2$) is larger by a factor of $\sim 10$ in
radius, or 1000 in volume. Likewise, the \ion{H}{2} region radius is
larger by a factor $> 10$ even compared to a GRB host with an extreme
Pop III population of stars. Furthermore, within the \ion{H}{2} region
itself, $\tau_{\rm res}$ is of order unity for a quasar, and of
order 10 in the GRB case. This means that absorption in this region
can be easily probed in quasars, while for GRB hosts the absorption is
near total except well into the damping wing; note that at larger
$\Delta D$, beyond the limits shown in the figure, $\tau_{\rm damp}$
continues to decrease roughly inversely with $\Delta D - \Delta D_{\rm
II}$, where $\Delta D_{\rm II} < 0$ is the apparent position of the
edge of the \ion{H}{2} region. Thus, cosmological infall is pronounced
around quasars and its effects are observable over a large region,
while it is weak in GRB hosts and essentially unobservable. On the
other hand, the damping wing can be observed in GRB hosts over a large
region to the red side of resonant absorption, while in quasars
$\tau_{\rm damp} \ll 1$ except in the region where $\tau_{\rm res}$
is of comparable importance.

Note that we have not solved for exact radiative transfer in the expanding
\ion{H}{2} region, but instead have assumed that the region is optically 
thin when we determined the ionizing intensity at various positions
within the region. In order to check this assumption, we calculate the
optical depth $\tau_{\rm ion}$ to hydrogen-ionizing photons from the
source being absorbed as they travel out to the edge of the \ion{H}{2}
region. This optical depth depends on the intensity and gas density at
all radii through the region. In order to place an upper limit on the
optical depth, we evaluate it for photons with an energy just above
the ionization threshold of hydrogen. For the three cases shown in the
figure, $\tau_{\rm ion}$ thus defined equals 0.019 (normal GRB host),
0.027 (Pop III GRB host) and 0.015 (quasar). The fact that $\tau_{\rm
ion} \ll 1$ justifies the neglect of radiative absorption and
scattering in our calculations.

\section{Detecting Cosmological Infall around Quasars}

As illustrated in the previous section, observations of \Lya
absorption in quasars can probe the cosmological infall pattern
induced by the dark matter halo surrounding each quasar. In this
section, we predict the dependence of the absorption pattern on the
quasar redshift and flux. If these trends predicted by theory can be
checked directly in observations, then this would test the fundamental
properties of quasars and of dark matter halos, as well as the
correlation between them.

Quasars have broad-band spectra that, due to their high luminosity,
can be measured accurately even at high redshift. However, unlike GRB
afterglows, the spectra are composed of emission lines broadened by
multiple components of moving gas, and the difficulty in predicting
the precise shape of, e.g., the \Lya line in any particular quasar
presents a significant systematic error in determinations of the
absorption profile. A promising approach is to adopt the parametrized
form of the emission profile that best fits the line shape of most
quasars at low redshift, and then fix the parameters for each quasar
based on the part of the \Lya line that does not suffer resonant
absorption \citep{nature}. Quasars also have important advantages. The
great intensity of quasars highly ionizes their host galaxies and
reduces the amounts of gas and dust, thus simplifying the
analysis. Furthermore, if the relatively high metallicity found for
high-redshift quasars is typical of quasars at even higher redshifts,
then various metal lines may be detected in addition to Ly$\alpha$,
allowing an independent measure of the host redshift and of the
velocity structure of the gas producing the broad emission lines.

We can easily understand the rough scaling with the observable
parameters (i.e., quasar luminosity and redshift) of the properties of
the red absorption drop due to infall. The luminosity of a quasar at
redshift $z_S$ is proportional to $M_{\rm BH}$ times $f_E$, the
fraction of the Eddington luminosity that is actually emitted by a
given quasar. We assume here a general power-law relation $M_{\rm BH}
\propto V_c^n$ between the black hole mass and the halo circular
velocity (note that we adopt $n=5$ in the numerical results, as
discussed in \S \ref{sec-IMFs}). The distance to the accretion shock
is proportional to the halo virial radius, and the infall velocity of
the pre-shock gas is proportional to the halo circular velocity. Thus,
the wavelength offset of the absorption drop (measured redward from
the central wavelength of the source \Lya emission) is \be
\Delta \lambda_{\rm drop} \propto (1+z_S)\, \left( \frac{L}{f_E}
\right)^{1/n} . \ee The halo mass is given by $M \propto 
(1+z_S)^{-3/2}\, V_c^3$, or \be M \propto (1+z_S)^{-3/2}\, \left(
\frac{L}{f_E} \right)^{3/n}\ . \ee The optical depth at the absorption 
drop is the Gunn-Peterson optical depth at that redshift, times the
relative gas density $\Delta$, times the neutral hydrogen
fraction. The neutral fraction is proportional to the gas density over
the ionizing intensity, where the intensity is proportional to the
quasar luminosity over the square of the distance to the accretion
radius.  Thus, the optical depth scales as \be \tau_{\rm drop} \propto
\Delta^2 (1+z_S)^{3/2}\, \frac{L^{(2-n)/n}} {f_E^{\,2/n}}\ , \ee where 
$\Delta$ depends on $M$ and $z$ through the infall model. Note that
the dependence of optical depth directly on halo mass is \be \tau_{\rm
drop} \propto \Delta^2 (1+z_S)^{(5-n)/2} f_E^{\,-1} M^{(2-n)/3}\ . \ee
Also note that clumping (\S~\ref{sec-infall}) modifies slightly these
scalings of the optical depth. Observing the absorption drops in
quasars over a wide range of luminosity and redshift will probe the
population of quasars and the properties of their halos through the
values of $f_E$, $\Delta$, and $n$.

\begin{figure}[htbp]
\plotone{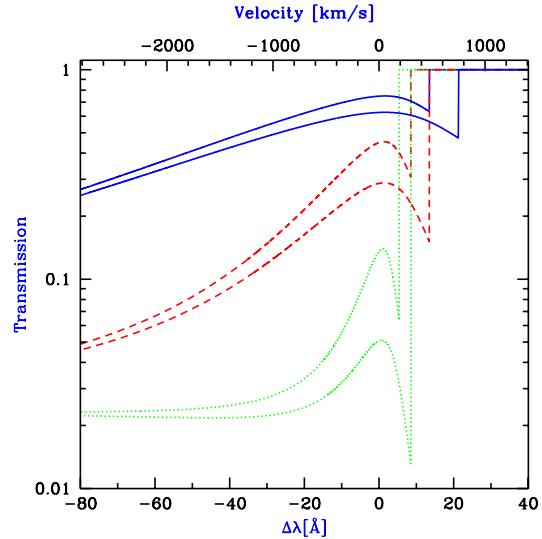} 
\caption{Transmission fraction of a quasar's flux through the 
infalling intergalactic medium, as a function of wavelength (or
velocity) relative to the center of the quasar's emission line. The
quasar is assumed to have an observed flux $F_{\lambda}$ of 1 (solid
curves), 0.1 (dashed curves), or 0.01 (dotted curves), in units of
$10^{-17}$ erg cm$^{-2}$ s$^{-1}$ \AA\ $^{-1}$. In each pair of
curves, the quasar is assumed to be radiating at its Eddington
luminosity (top) or at $10\%$ of Eddington (bottom). All curves are at
redshift 6 and assume that the quasar had turned on after
reionization.}
\label{fig-QSO1}
\end{figure}

Figure~\ref{fig-QSO1} shows the fraction of the quasar's flux that is
transmitted through the surrounding, infalling IGM. We assume a quasar
at redshift 6 that had turned on after reionization, and we therefore
include only resonant absorption. We consider a quasar with a flux
typical of the recent SDSS high-redshift quasars, and also compare to
quasars with fluxes lower by a factor of 10 or 100, applicable to
future surveys. In each case, we translate the flux to black hole mass
by assuming that the quasar radiates at the Eddington luminosity. We
also compare to the case where the quasar is instead assumed to be
radiating at only $10\%$ of Eddington; in this case the black hole
must be more massive in order to produce a given observed flux,
implying a larger surrounding halo, and an absorption drop more
strongly shifted to the red (corresponding to a larger infall
velocity). As implied by the scalings presented above, fainter quasars
at a given redshift have a smaller incursion of absorption into the
red wing of Ly$\alpha$; they also have a larger $\tau_{\rm drop}$
since the slightly smaller $\Delta$ around lower-mass hosts does not
compensate for the lower luminosity of the quasars.

\begin{figure}[htbp]
\plotone{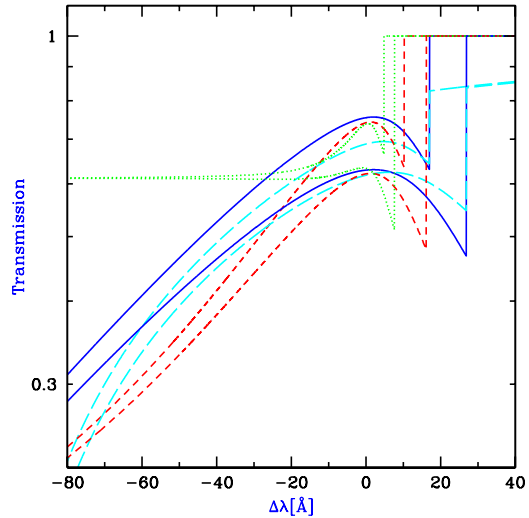}
\caption{Transmission fraction of a quasar's flux through the
infalling intergalactic medium, as a function of wavelength relative
to the center of the quasar's emission line. The quasar is assumed to
have an observed flux $F_{\lambda}$ of $10^{-17}$ erg cm$^{-2}$
s$^{-1}$ \AA\ $^{-1}$, and to be in a post-reionization universe at
redshift 7 (solid curves), 5 (short-dashed curves), or 3 (dotted
curves). Also shown is the same redshift 7 quasar but assuming that it
reionized its surroundings for the first time over a time $t_S=2
\times 10^7$ yr (long-dashed curves). In each pair of curves, the
quasar is assumed to be radiating at its Eddington luminosity (top) or
at $10\%$ of Eddington (bottom).}
\label{fig-QSO2}
\end{figure}

The transmitted flux fraction is also shown in Figure~\ref{fig-QSO2},
but for various quasar redshifts. For a given observed flux,
lower-redshift quasars are situated inside smaller hosts, and the
absorption profiles are seen to have a smaller $\Delta \lambda_{\rm
drop}$ along with a slightly lower $\tau_{\rm drop}$. For $z_S=7$, a
quasar emitting inside a still neutral region is distinguished by the
appearance of the damping wing. In addition, such a quasar has a
significantly smaller red absorption drop because of the higher
temperature of gas reionized by the relatively hard quasar spectrum;
the recombination rate is lower in such gas, thus reducing the
hydrogen neutral fraction and the \Lya optical depth.

In both figures, apparent at large distances from the quasar is the
contribution of the mean ionizing background, which is included at
redshift 6 and below but is most prominent at redshift 3. In addition,
the appearance of the red-most absorption feature, and the shape of
the rise and then the fall of the transmission toward shorter
wavelengths, vary widely depending on the properties of the quasar,
its halo, and the surrounding infall region. In principle, the precise
shape of the absorption profile can be used to measure the detailed
gas density profile in addition to the basic parameters of the quasar
and its host halo. However, these transmission profiles cannot be
directly observed; for any particular quasar, the transmission profile
operates on the intrinsic profile of the quasar emission, and density
fluctuations along the line of sight change the transmission relative
to the ensemble-averaged profile. While these complications can be
modeled, we have focused here on the observable features that are
independent of these uncertainties and that are recognizable even in
individual spectra, although the precise values may fluctuate compared
to the ensemble-averaged ones; specifically, these features are the
magnitude of the red absorption drop, and its location in wavelength,
which is most useful if the quasar redshift is measured accurately,
e.g., from other emission lines. For a statistical sample of quasars,
our results should best match the ensemble-averaged spectrum.

\section{Probing Reionization with $\gamma$-ray Burst Afterglows}

\label{sec-GRBs}

Gamma-Ray Burst (GRB) explosions from the first generation of stars
offer a particularly promising opportunity to probe the epoch of
reionization.  Young (days to weeks old) GRBs are known to outshine
their host galaxies in the optical regime at $z\sim 1-3$. In
hierarchical galaxy formation in $\Lambda$CDM, the characteristic mass
and hence optical luminosity of galaxies and quasars declines with
increasing redshift; hence, GRBs should become easier to observe than
galaxies or quasars at increasing redshift. If the first stars are
preferentially massive \citep{Abel, BCL2002} and GRBs originate during
the formation of compact stellar remnants such as neutron stars and
black holes, then the fraction of all stars that end up as GRB
progenitors may be even higher at early cosmic times.

Similarly to quasars, GRB afterglows possess broad-band spectra which
extend into the rest-frame UV and can probe the ionization state and
metallicity of the IGM out to the epoch when it was reionized
\citep{LR00}. Simple scaling of the long-wavelength spectra and
temporal evolution of afterglows with redshift implies that at a fixed
time lag after the GRB trigger in the observer's frame, there is only
a mild change in the {\it observed}\, flux at infrared or radio
wavelengths as the GRB redshift increases \citep{Ciardi}. Hence, GRBs
provide exceptional lighthouses for probing the universe at high
redshift. The infrared spectrum of GRB afterglows occurring around the
reionization redshift could be taken with future telescopes such as
the James Webb Space Telescope ({\it JWST}; see
http://www.jwst.nasa.gov/), as a follow-up on their early X-ray
localization with the {\it Swift}\, satellite. In this section, we
explore how such a spectrum could reveal the damping wing due to
neutral IGM, and thus probe reionization.

If GRBs originate from stellar remnants of high-mass stars, then they
are expected to occur in host galaxies with a probability proportional
to the SFR of each galaxy. In order to specify a typical host, we
estimate as follows the SFR of galaxies in halos of various mass $M$
at redshift $z=7$. We assume that the SFR in a given halo is
proportional to $M$ divided by a star-formation timescale which is
determined by the halo merger history. Specifically, we estimate the
age of gas in a given halo using the average rate of mergers which
built up the halo. Based on the extended Press-Schechter formalism
\citep{lc93}, for a halo of mass $M$ at redshift $z$, the fraction of
the halo mass which by some higher redshift $z_2$ had already
accumulated in halos with galaxies is \be F_M(z,z_2) = {\rm erfc}
\left(\frac{1.69/D(z_2)- 1.69/D(z)}{\sqrt{2 (S(M_{\rm
min}(z_2))-S(M))}} \right)\ , \ee where $D(z)$ is the linear growth
factor at redshift $z$, $S(M)$ is the variance on mass scale $M$
(defined using the linearly-extrapolated power spectrum at $z=0$), and
$M_{\rm min}(z_2)$ is the minimum halo mass for hosting a galaxy at
$z_2$. We assume that prior to the final phase of reionization this
minimum mass is determined by the minimum virial temperature of $\sim
10^4$ K required for efficient atomic cooling in gas of primordial
composition. We estimate the typical age of gas in the halo as the
time since redshift $z_2$ where $F_M(z,z_2) = 0.5$, so that $50\%$ of
the gas in the halo has fallen into galaxies only since then. The
result at $z=7$ is that $M_{\rm min}(z=7)=1.3 \times 10^8 M_{\odot}$,
$50\%$ of the total SFR occurs in halos up to $M=4 \times 10^8
M_{\odot}$, and $90\%$ in halos up to $M=5 \times 10^9 M_{\odot}$.

The GRB afterglow spectrum can be modeled as a set of power-law
segments. Regardless of the particular power law which applies to the
\Lya region, since the wavelength band relevant for \ion{H}{1}
absorption is narrow, the intrinsic afterglow emission is essentially
constant over this region; e.g., at $z=7$, the flux level should vary
by $\la 1\%$ over a range of $\Delta \lambda=100$\AA\ around the
redshifted \Lya wavelength. Thus, the GRB afterglow offers a unique
opportunity to detect the signature of IGM absorption, if the
afterglow spectrum is at all similar to the theoretically expected
smooth spectrum. The following numerical expressions and figures are
based on the model of \citet{sari1}, corrected for a jet geometry
using \citet{sari2} [see also \citet{EliGRB}].

Based on modeling of observed afterglows, we assume typical afterglow
parameters of $E=10^{51}$ erg for the total (not isotropic) energy
\citep{F01}, a jet opening angle of $\theta=0.07$ rad, an external
shock occurring in a surrounding medium of uniform density $n=1$
cm$^{-3}$, electrons accelerated to a power-law distribution of
Lorentz factors with index $p=2.4$, and a fraction $\epsilon_B=10\%$
of the shock energy going to the magnetic field and $\epsilon_e=10\%$
to the accelerated electrons. We assume adiabatic evolution, and
neglect synchrotron self-absorption which is only important at much
lower frequencies than we consider.

\begin{figure}[htbp]
\plotone{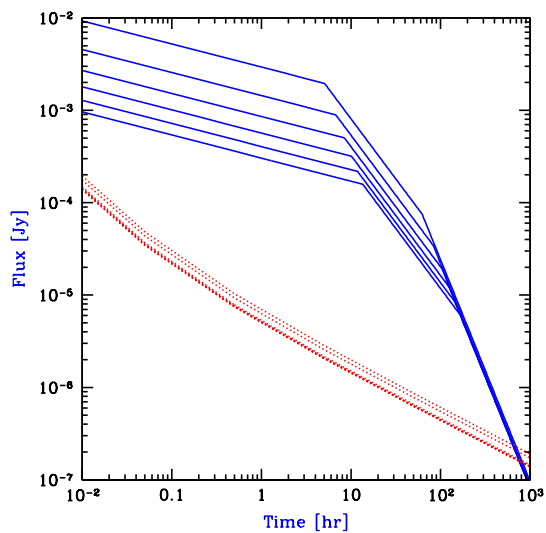} 
\caption{Detectability of high-redshift GRB afterglows as a function
of time since the GRB explosion as measured by the observer. The GRB
afterglow flux is shown (solid curves) at the redshifted \Lya
wavelength. Also shown (dotted curves) is the detection threshold for
{\it JWST}\, assuming a spectral resolution $R=5000$ with the near
infrared spectrometer, a signal to noise ratio of 5 per spectral
resolution element, and an exposure time equal to $20\%$ of the time
since the GRB explosion. In each set of curves, a sequence of
redshifts is used, $z=5$, 7, 9, 11, 13, and 15, respectively, from top
to bottom.}
\label{fig-GRB1}
\end{figure}

Figure~\ref{fig-GRB1}\footnote{Detection thresholds in
Figure~\ref{fig-GRB1} were obtained using the {\it JWST}\, calculator
at http://www.stsci.edu/jwst/science/jms/index.html .} shows
quantitatively the detectability of GRB afterglows at redshifts
$z=5$--15. The figure indicates that the redshifted \Lya region of GRB
afterglows remains sufficiently bright for a high-resolution study
with {\it JWST}\, up to several weeks after the GRB explosion,
throughout the redshift range we consider. Each light-curve of
afterglow flux shows a couple of breaks over the range of times
considered in the figure. We let $\nu_m$ and $\nu_c$ be the
characteristic synchrotron frequencies (redshifted to the observer)
of, respectively, the lowest-energy electron in the accelerated
power-law distribution, and the lowest-energy electron that radiates
efficiently enough to lose most of its energy. At early times the
afterglow is in the fast cooling regime ($\nu_m > \nu_c$), when all
the accelerated electrons cool efficiently, and the observed frequency
$\nu_{\alpha}$ (corresponding to redshifted Ly$\alpha$) is
intermediate between the two characteristic frequencies, leading to a
spectral shape $F_{\nu} \propto \nu^{-0.5}$ near \Ly. The first break
occurs when $\nu_{\alpha}$ rises above $\nu_m$ at \ba t_{\rm obs} & &
= 6.8\, \left(\frac{E}{10^{51}\,{\rm erg}} \right)^ {1/3}\,
\left(\frac{\theta}{0.07\,{\rm rad}} \right)^{-2/3}\,
\left(\frac{\epsilon_B}{0.1}\right)^{1/3} \nonumber \\ & & \times 
\left(\frac{\epsilon_e} {0.1}\right)^{4/3}\, \left(\frac{1+z_S}{8} 
\right)\, {\rm hrs}\ . \ea After this break, $\nu_{\alpha}$ is bigger 
than both $\nu_m$ and $\nu_c$, and $F_{\nu} \propto \nu^{-p/2}$ near
\Ly. The second break, which is due to sideways expansion of the jet,
affects the light-curve but not the spectral slope near Ly$\alpha$. It
occurs at \ba t_{\rm obs} & & = 3.5\, \left(\frac{E}{10^{51}\,{\rm
erg}} \right)^ {1/3}\, \left(\frac{\theta}{0.07\,{\rm rad}} \right)^2
\nonumber \\ & & \times \left(\frac{n}{1\,{\rm cm}^{-3}}\right)^{-1/3}\, 
\left(\frac{1+z_S}{8} \right)\, {\rm days}\ . \ea

As noted earlier, afterglows fade only slowly with redshift, since a
given observed time corresponds to an earlier source time for high
redshift afterglows, which are thus intrinsically brighter. Indeed,
after the second break, the \Lya flux at a given $t_{\rm obs}$ is
enhanced by this effect in proportion to $(1+z_S)$ to the power $p$, a
factor that for $p=2.4$ almost overcomes the fading according to the
inverse squared luminosity distance. GRB afterglows may be even more
easily detectable than indicated in figure~\ref{fig-GRB1} if a
significant fraction exhibit an early UV-optical flash from a reverse
external shock; e.g., the flash in GRB 990123 at $z = 1.6$ was
detected at a flux level of $\sim 1$ Jy at rest-frame $\lambda \sim
2000$\AA\ about 50 seconds after the burst began \citep{GRBflash}.

Relative to other possible sources, GRB afterglows offer the advantage
of having smooth, power-law spectra, which helps GRBs avoid the
systematic uncertainties associated with the need to model the
\Lya emission profiles of galaxies and quasars. In addition, GRBs are 
expected to occur mainly in relatively low mass halos, which induce
only weak infall of the surrounding gas and produce insignificant
\ion{H}{2} regions, as illustrated in \S~\ref{sec-illust}. Metal lines 
from the host may be seen in absorption in the afterglow spectrum,
thus allowing an accurate redshift measurement. Emission from the host
galaxy should not significantly contaminate the early afterglow, but
it may make the galaxy marginally detectable with {\it JWST}\, once
the afterglow has faded away. For instance, $10\%$ of the SFR at $z=7$
occurs in halos above $M=5 \times 10^9 M_{\odot}$, and a halo of this
mass is expected to have a SFR of $\sim 1 M_{\odot}$ per year, which
translates [for a normal IMF] to a 1500\AA\ continuum flux density of
$\sim 15$ nJy, and a flux density in the $\sim 10$\AA\ bandwidth of
the \Lya line emission from the galaxy of $\sim 500$ nJy. If the
star-forming regions where GRBs are expected to occur contain
significant column densities of gas and dust, then they may
contaminate the signal from IGM absorption, as discussed in
\S~\ref{sec-extinct}. 

\begin{figure}[htbp]
\plotone{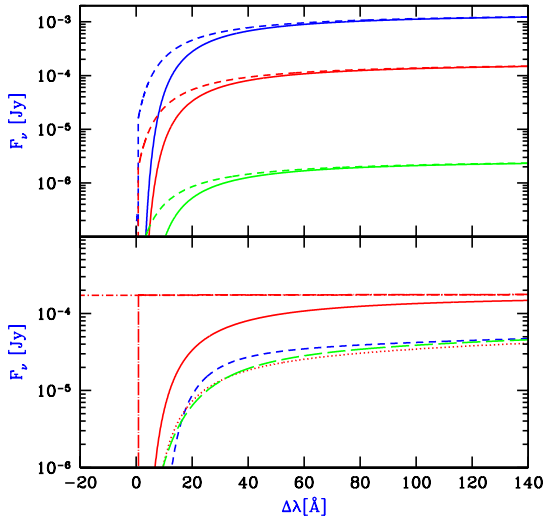} 
\caption{IGM absorption profiles in GRB afterglows, presented in terms
of the flux density $F_{\nu}$ versus relative observed wavelength
$\Delta \lambda$. All curves adopt $z=7$ (assumed prior to the final
reionization phase) and the typical halo mass $M=4 \times 10^8
M_{\odot}$ expected for GRB host halos. Top panel: Predicted spectrum
including IGM \ion{H}{1} absorption (both resonant and damping wing),
for host galaxies with $t_S=10^7$ yr, $f_{\rm esc}=10\%$ and a normal
IMF (solid curves) or $t_S=10^8$ yr, $f_{\rm esc}=90\%$ and a Pop III
IMF (dashed curves). The observed time after the burst is one hour,
one day, and ten days, from top to bottom, respectively. Bottom panel:
Predicted spectra one day after a GRB for a host galaxy with
$t_S=10^7$ yr, $f_{\rm esc}=10\%$ and a normal IMF. Shown is the
unabsorbed GRB afterglow (dot-short dashed curve, essentially
horizontal), the afterglow with resonant IGM absorption only (dot-long
dashed curve), and the afterglow with full (resonant and damping wing)
IGM absorption (solid curve). Also shown, with 1.7 magnitudes of
extinction, are the afterglow with full IGM absorption (dotted curve),
and attempts to fit this profile with a damped \Lya absorption system
in the host galaxy (dashed curves; see text).}
\label{fig-GRB2}
\end{figure}

Figure~\ref{fig-GRB2} illustrates GRB afterglow absorption profiles
(top panel) and explores the possibility of contamination from dust
and neutral gas in the host (bottom panel). If absorption from the
host is negligible then GRB afterglows offer a clear and direct
measurement of absorption by the damping wing of \ion{H}{1} in the
IGM. In particular, the sharp cutoff expected if only resonant
absorption is included (bottom panel, dot-long dashed curve) is easily
distinguishable from the gradual cutoff characteristic of absorption
by the damping wing (bottom panel, solid curve). In this case, the IMF
can be constrained as well, since a strongly-ionizing IMF produces a
gradual cutoff as well as a sharp drop due to resonant absorption (top
panel, dashed curves); this occurs since the neutral gas lies further
away (in the blue wing), producing a relatively weak damping wing that
is interrupted by the ubiquitous red-wing drop due to resonant
absorption.

However, as illustrated in the bottom panel of figure~\ref{fig-GRB2},
contamination by absorption in the host galaxy may hamper studies of
reionization and of the IMF based on absorption profiles. The three
lower curves include extinction as might be produced by an \ion{H}{1}
column density of $10^{21}\,{\rm cm}^{-2}$ [see
eq.~(\ref{eq-extinc})]. Because of its relatively smooth spectrum,
extinction does not significantly change the shape of the absorption
profile; moreover, it is possible to take advantage of the broad-band
afterglow spectrum in order to derive the effect of extinction near
redshifted \Lya by measuring the extinction over a much wider range of
wavelengths than shown in the figure and interpolating into the
critical bandpass. However, extinction does lower the overall flux and
thus increases the integration time required to measure an accurate
spectrum. 

As shown in figure~\ref{fig-GRB1}, at $t_{\rm obs}=1$ day, a minimum
flux of $F_{\nu} \sim 10^{-6}$ Jy is needed for a high-resolution
spectrum with {\it JWST}.\, Assuming a spectrum obtained with this
sensitivity, we consider whether absorption by a large \ion{H}{1}
column density in the host could be confused with the signature of the
IGM damping wing before reionization. We consider only the $F_{\nu} >
10^{-6}$ Jy portion of the profile with extinction and full absorption
due to the IGM (figure~\ref{fig-GRB2}, bottom panel, dotted curve),
and attempt to reproduce this profile by replacing the IGM damping
wing with a damped \Lya system in the galactic host. If the host
redshift is precisely known, the different shapes of the two gradual
cutoffs preclude any possibility of confusion between them. For
example, a damped \Lya system with $N_{\rm H\, I}=2\times
10^{20}\,{\rm cm}^{-2}$ (short-dashed curve) is 1.5 orders of
magnitude too low at $\Delta \lambda \sim 10$\AA\ and 10--20$\%$ too
high at $\Delta \lambda = 30$--100\AA.

However, if the source redshift is not measured exactly then a much
closer match is possible. The damped \Lya system (long-dashed curve)
that most closely simulates the effect of the IGM damping wing in this
case has $N_{\rm H\, I}=7.7\times 10^{20}\,{\rm cm}^{-2}$ and is
shifted blueward by $\Delta \lambda = -15.4$\AA, equivalent to $\Delta
z$ = 0.013 or a velocity blueshift of 500 km s$^{-1}$. This match,
although it appears fairly close in the figure, is off at various
wavelengths by a difference ranging between $-6\%$ and $+6\%$, and
this difference should be easily measurable. However, the relatively
close match in this comparison of a pure IGM damping wing with a pure
damped \Lya system suggest that in a realistic situation, when both
may contribute, this partial degeneracy will limit the constraints
that can be derived on the various parameters of interest. Note,
though, that since the halo circular velocity is only $\sim 24$ km
s$^{-1}$, the peculiar velocity of the GRB (which may arise from the
circular velocity of the disk) is negligible, and an accurate redshift
measurement can help to break the near-degeneracy. Regarding another
possible difficulty, \citet{jordi98} showed that $\sim 10\%$
uncertainties in the cosmological parameters would contribute further
to the near-degeneracy between a damped \Lya system and the IGM
damping wing, but this factor continually becomes less important as
various cosmological probes yield increasingly precise parameter
values \citep[e.g.,][]{WMAP}.

The hydrogen absorption within the host galaxy of the GRB should be
reduced by the early ionizing radiation of the GRB afterglow itself
\citep{Perna98}. The recombination time of the surrounding
interstellar medium ($\sim 10^5 n_0^{-1}~{\rm yr}$) is long compared
to the duration of the afterglow emission, where $n_0$ is the medium
density in units of ${\rm cm^{-3}}$. Hence, each ionizing photon will
eliminate a neutral hydrogen atom from the interstellar medium along
the line of sight. Given that the number of hydrogen atoms out to a
distance $r$ is $\propto n_0 r^3$, the radius ionized by a given
afterglow varies as $n_0^{-1/3}$. \citet{DrH02} showed that an early
optical/UV flash should ionize hydrogen and destroy dust and molecular
hydrogen out to a distance of $\sim 10^{19}$ cm in surroundings of
$n_0 \sim 10^3-10^4$ \citep[see also][]{Perna98,WDr00, Dr00,
Fr01,Perna02,Perna03}. Therefore, the early afterglow emission could
significantly reduce the damped Ly$\alpha$ absorption within a host
molecular cloud ($n_0\sim 10^4$, $r\sim 10$ pc) or a line of sight
passing at an angle through a host galactic disk ($n_0\sim 1$, $r\sim
100$ pc).  However, the afterglow's characteristic sphere of influence
is negligible compared to the intergalactic scales of interest for
reionizing the IGM.

\section{Conclusions}

We have calculated the profile of Ly$\alpha$ absorption for GRBs and
quasars during and after the epoch of reionization, including the
effects of ionization by the source and its host galaxy, clumping, and
cosmological infall of the surrounding cosmic gas. We find that GRBs
are the optimal probes of the ionization state of the surrounding IGM
as they are most likely to reside in dwarf galaxies that do not
significantly perturb their cosmic environment. On the other hand,
bright quasars reside in the most massive galaxies and probe the
strong gravitational effects of these galaxies and their halos on the
IGM.

In order to isolate the intergalactic contribution to the Ly$\alpha$
damping wing, it is necessary to measure the source redshift
accurately. Here again, GRB afterglows are simpler to interpret. In
quasars, although there are often strong emission lines in addition to
Ly$\alpha$, bulk motions can shift them substantially from the
systemic redshift of the host galaxy. For example, some broad emission
lines such as \ion{C}{4} are typically offset by $\sim 1000$ km
s$^{-1}$, but low-ionization broad lines such as \ion{Mg}{2} are only
offset by $\sim 200$ km s$^{-1}$ \citep{Richards}. The \ion{Mg}{2}
line can be detected at high redshift in the near-infrared and,
indeed, it has been observed for the highest redshift quasar known
\citep{Willott}. Redshifts accurate to better than 100 km
s$^{-1}$ may be obtainable by detecting narrow lines such as
[\ion{O}{3}], perhaps marginally possible from the ground or else
doable with {\it JWST}.\, For either a GRB or a quasar, however,
perhaps the best method for determining its redshift is to detect
molecular emission or absorption lines in the interstellar medium of
its host galaxy. The resulting measurement is likely to be accurate to
within the disk velocity dispersion, which should be rather low,
particularly in the low-mass hosts that are typical of GRBs. These
lines can be detected at high redshift, as demonstrated by the CO
emission from the host galaxy of the highest-redshift quasar known
\citep{COhighz}. Note that another possible disadvantage of quasars is
that their massive hosts may develop significant peculiar velocities
as they are likely to reside in rare high-density environments such as
groups of galaxies.

Supernova and quasar outflows could change our predictions if they
reached out from the galaxy far enough to affect the gas at the
distances where it produces the absorption in our model. Although
outflowing gas can be detected spectrally and the outflow velocity
deduced, it is generally very difficult to determine the spatial scale
of an outflow around a high-redshift galaxy. \citet{Adelberger} have
detected a statistical signature based on six objects that suggests
that Lyman break galaxies at $z \sim 3$--4 produce outflows that drive
away surrounding gas out to a radius of $\sim 0.5\ h^{-1}$ comoving
Mpc. If this preliminary signal is correct, and if these Lyman break
galaxies lie in $M \ga 10^{12} M_{\odot}$ host halos, then these
outflows reach a distance comparable to the halo accretion shock and
may begin to marginally affect our model predictions. Note that direct
absorption by the outflowing gas should generally be distinguishable
from the absorption patterns that we predict. Absorption from outflows
should occur in the blue wing and thus not interfere with the red
absorption feature due to infall. Furthermore, outflowing gas is
likely to be highly enriched and can be probed (or ruled out) in each
object using metal lines.

High-redshift GRB afterglows are likely to open a new frontier in
extragalactic astronomy (``{\it GRB cosmology}''), by providing
important new clues about their progenitors --- the first stars ---
and about the reionization history of the intervening IGM.  Afterglow
spectra can also be used to explore the metal enrichment history of
the IGM through the detection of intergalactic metal absorption lines
\citep{fl03}. The comparison between observed spectra of high-redshift
afterglows and detailed cosmological simulations of the IGM (involving
hydrodynamics and radiative transfer during the epoch of reionization)
will provide a better understanding of the process of galaxy formation
in the young universe, only $\sim 10^8$--$10^{9}$ years after the big
bang.

\acknowledgments

We acknowledge joint support by NSF grant AST-0204514 and NATO grant
PST.CLG.979414. R.B. is grateful for the support of an Alon Fellowship
at Tel Aviv University and of Israel Science Foundation grant 28/02.
A.L. acknowledges support from the Institute for Advanced Study at
Princeton and the John Simon Guggenheim Memorial Fellowship. This work
was also supported in part by NSF grant AST-0071019 and NASA grant
ATP02-0004-0093 (for A.L.).

\end{document}